\documentclass[aps,prb,preprint,amsfonts,superscriptaddress,floatfix]{revtex4-2}
\usepackage{graphicx}% Include figure files
\usepackage{dcolumn}% Align table columns on decimal point
\usepackage{bm}% bold math
\usepackage[utf8]{inputenc}
\usepackage[T1]{fontenc}
\usepackage{mathptmx}
\usepackage{comment}

\usepackage{pdfpages} % include pdfs
\usepackage{pgffor} % for loops

\makeatletter
\AtBeginDocument{\let\LS@rot\@undefined}
\makeatother

\def\supplementfilename{SM.pdf}

\pdfximage{\supplementfilename}
\def\numbersupplementpages{\the\pdflastximagepages}

\newif\ifarXiv
\arXivtrue

\begin{document}

\preprint{AIP/123-QED}

\title{Magnon-magnon coupling efficiency of $\eta$=0.5 in weakly pinned synthetic antiferromagnets}

\author{Dirk Backes}%
 \email{dirk.backes@diamond.ac.uk}
\affiliation{Diamond Light Source, Harwell Campus, Didcot OX11 0DE, United Kingdom}

\date{\today}% It is always \today, today,
             %  but any date may be explicitly specified

\begin{abstract}
Synthetic antiferromagnets (SAFs) consist of two ferromagnetic layers that are antiferromagnetically coupled. These systems support complex dynamical magnetic excitations, where interlayer coupling gives rise to both in-phase (acoustic) and anti-phase (optical) magnonic modes. Typically, simultaneous excitation of both modes requires breaking the symmetry between the ferromagnetic layers -- commonly achieved through slight misalignment of the experimental setup or by modifying the intrinsic magnetic properties.
In our approach, we utilize a pinned synthetic antiferromagnet (pSAF), where one of the ferromagnetic layers is exchange-coupled to an antiferromagnet. We demonstrate that by tuning the thickness of the antiferromagnet and slightly enhancing the magnetic anisotropy of the pinned layer, both acoustic and optical modes can be efficiently excited -- without the need for experimental misalignment or changes to the intrinsic material properties.
Under specific conditions, the magnon dispersion relations exhibit anti-crossing behavior, resulting in the emergence of a magnonic bandgap -- a clear signature of strong magnon-magnon coupling. The coupling efficiency $\eta$, defined as the ratio between the bandgap and the characteristic frequency, reaches $\eta = 0.5$, approaching the ultra-strong coupling regime ($\eta = 1$).
The combination of strong mode hybridization, a sizable magnonic bandgap, and high ferromagnetic resonance (FMR) coherence over large areas -- all achieved at room temperature without cryogenic cooling -- underscores the potential of these systems for quantum magnonic applications, including quantum computing.
\end{abstract}
	
\maketitle

%\section{\label{sec:level1}Introduction}

Hybridization between quantum systems and electromagnetic fields -- typically observed as avoided level crossings -- lies at the heart of many quantum technologies. In superconducting qubit platforms, for example, strong coupling to microwave cavities enables coherent control and read-out\,\cite{Majer2007}. However, these systems face fundamental challenges, including limited coherence times, sensitivity to noise, and the need for cryogenic environments. As a result, there is growing interest in alternative platforms that offer scalable architectures, improved stability, and operation at or near room temperature.

Magnonic systems based on ferro- or antiferromagnetic layers have emerged as promising platforms for quantum computation. They offer key advantages, including operation at room temperature and highly tunable resonance frequencies via external magnetic fields\,\cite{Yuan2022,Jiang2023}. Coupling between magnons and photons has already been demonstrated\cite{Li2019a, Hou2019, Lee2023}, and intriguingly, a two-level system -- a fundamental building block of quantum computing -- can also be realized by coupling magnonic modes to each other.
Synthetic antiferromagnets (SAFs)\,\cite{VanDenBerg1996} are particularly attractive in this context, as they support a rich spectrum of magnonic modes in the GHz frequency range, accessible at moderate magnetic fields (<\,1\,T) and without the need for cryogenic cooling. SAFs are magnetically compensated due to the antiferromagnetic interlayer coupling, resulting in negligible net magnetization. This interlayer exchange couples the magnetization precession in the individual ferromagnetic layers, giving rise to two fundamental collective modes: the acoustic mode, where the layers precess in phase, and the optical mode, where they precess with a 180° phase difference\,\cite{Liu2014, Tanaka2014, Chiba2015, Timopheev2014, Waring2020, Shiota2020, Sud2020, Li2021, Li2023, Dai2021}.

When the acoustic and optical dispersions intersect, the two magnon modes can hybridize. This magnon-magnon coupling (MMC) manifests as a gap opening between the upper and lower branches of the dispersion spectrum -- analogous to avoided level crossings in coupled two-level systems\,\cite{Waring2020, Shiota2020, Sud2020}. MMC has been observed not only in SAFs, but also in van der Waals magnets\,\cite{Macneill2019, Sklenar2021}, ferrimagnets\,\cite{Liensberger2019}, and patterned magnonic media\,\cite{Chen2018a, Adhikari2020}.
Two main challenges must be addressed for practical implementations:
Excitation of the optical mode is nontrivial due to its zero net dynamic magnetic moment, which makes it weakly responsive to uniform ac magnetic fields.
Establishing an effective mixing mechanism is essential to enable strong coupling between the optical and acoustic modes, thereby forming hybridized magnonic states suitable for quantum information processing.
  
Previous studies have explored various strategies to enable excitation and coupling of magnon modes in SAFs. These include intentional misalignment of the sample relative to the applied magnetic field\,\cite{Sud2020}, the use of specially designed RF waveguides to generate non-uniform excitation fields\,\cite{Shiota2020}, and the implementation of intrinsically imbalanced magnetic structures such as synthetic ferrimagnets\,\cite{Sud2023, Ma2023}. However, each of these approaches presents notable drawbacks: they are either technically complex or compromise the magnetically compensated nature of the SAF -- an essential property that minimizes stray magnetic fields and is critical for many applications.

In this Letter, we demonstrate a synthetic antiferromagnet design with intrinsic asymmetry that preserves the magnetically compensated nature of the system. The core idea is to introduce this asymmetry by coupling one of the two ferromagnetic layers to an antiferromagnet, such as PtMn. By varying the PtMn thickness, the strength of the exchange pinning is tuned to enhance the visibility of the optical magnon mode and simultaneously open a pronounced magnonic band gap at the intersection of the acoustic and optical modes. The resulting band gap exceeds all previously reported values in SAF-based systems, underscoring the potential of this approach as a powerful platform for magnonics-based quantum and classical information processing.

%\section{Sample preparation}

\begin{figure}
\includegraphics[width=10cm]{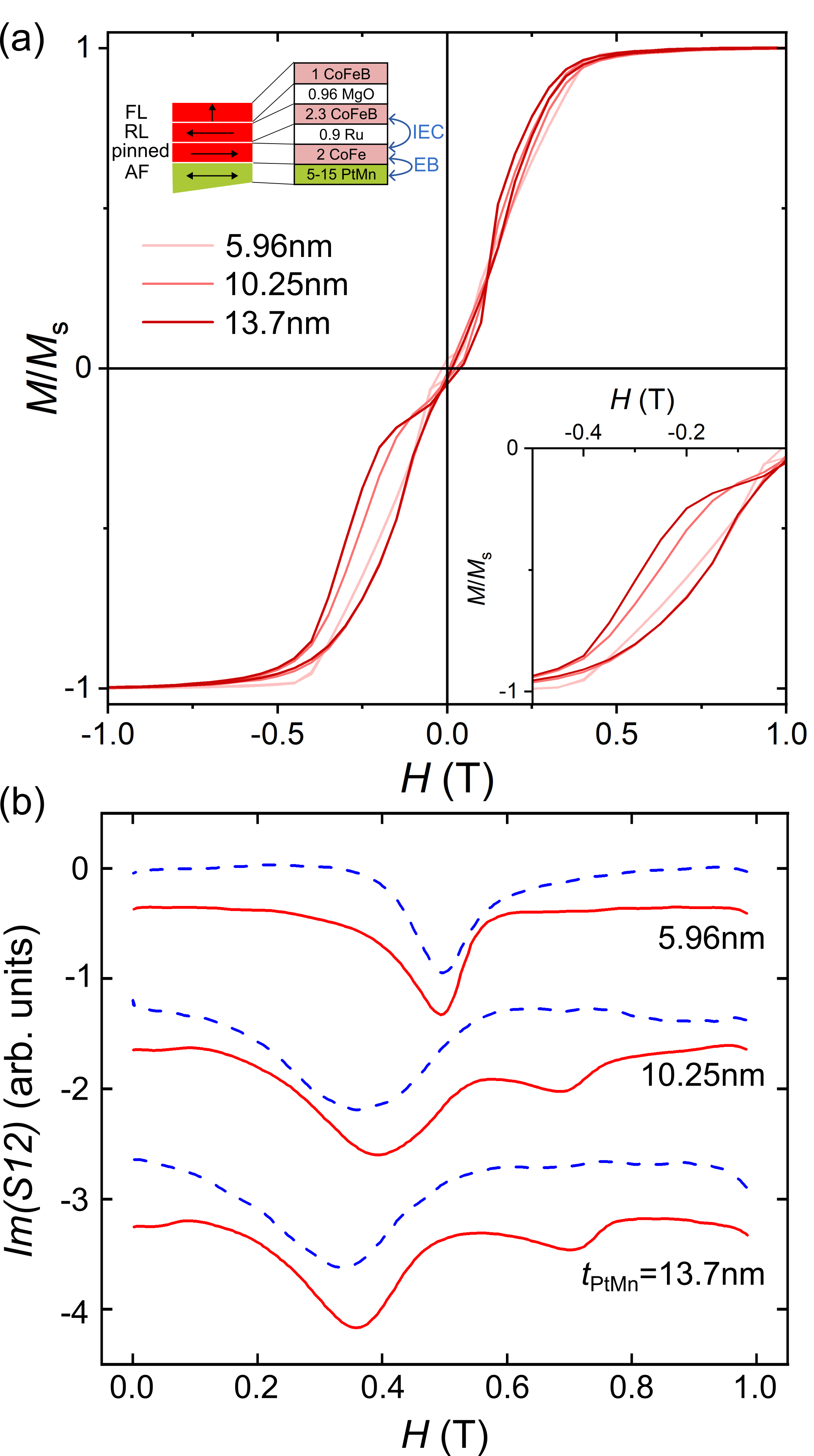}
\caption{\label{fig1} (a) VSM hysteresis loops of the pSAF with 3 different PtMn thicknesses. The schematic shows the material, thickness and function of the layers of the stack. (b) FMR spectra of the pSAF sample (straight) and a bilayer of PtMn and CoFeB (dashed lines) with the same PtMn thicknesses as in (a).}
\end{figure}

Spin-valve layer stacks with the composition of 3 Ta/40 Cu(N)/5 Ta/5-15 Pt$_{38}$Mn$_{62}$/2 Co$_{70}$Fe$_{30}$/\allowbreak0.9 Ru/2.3 Co$_{40}$Fe$_{40}$B$_{20}$/0.96 MgO/1 Co$_{40}$Fe$_{40}$B$_{20}$/3 Ta/7 Ru (thicknesses in nm) were deposited by sputtering on Si substrates with a 100\,nm SiO$_x$ layer. After deposition, the samples were annealed in a 1\,T magnetic field and subsequently field-cooled at 300$^\circ$\,C for another hour to set the exchange bias along a defined in-plane direction. Figure\,\ref{fig1}(a) shows the main functional layers of the stack. The ferromagnetic layer exchange-coupled to the antiferromagnetic PtMn layer acts as the pinned layer. It is separated from a second ferromagnetic layer -- the reference layer (RL) -- by a thin Ru spacer, forming a pinned synthetic antiferromagnet (pSAF). Adjacent to the reference layer lies a free layer, separated by an MgO tunnel barrier. For FMR analysis, it is desirable that the resonance peak associated to the free layer does not appear in the in-plane spectra. Removing the free layer entirely would, however, likely disrupt the crystalline structure of the full stack. To preserve structural integrity while suppressing the free-layer signal, the thickness of the free layer was reduced to enhance its perpendicular magnetic anisotropy\,\cite{Ikeda2010}, effectively shifting its resonance out of the in-plane spectra. As a control, a simplified stack of 3 Ta/40 Cu(N)/5 Ta/5-15 PtMn$_\mathrm{62}$/2 Co$_\mathrm{70}$Fe$_\mathrm{30}$/7 Ru was grown, omitting both the free and reference layers and retaining only the exchange bias bilayer.

The pinned and reference layers in the pSAF are coupled antiparallel via the Ruderman–Kittel–\allowbreak Kasuya–Yosida (RKKY) interaction. The Ru spacer thickness, $t_\mathrm{Ru}$, was optimized to achieve maximum antiferromagnetic coupling. To ensure full magnetic compensation of the pSAF, the thicknesses of the pinned and reference layers were carefully adjusted to satisfy the condition $t_\mathrm{pin} \times M_\mathrm{s,pin} = t_\mathrm{RL} \times M_\mathrm{s,RL}$ where $t_\mathrm{pin}$ and $t_\mathrm{RL}$ are the respective layer thicknesses, and $M_\mathrm{s,pin}$ and $M_\mathrm{s,RL}$ are the saturation magnetizations. In this design, $t_\mathrm{pin}$ is additionally optimized to ensure strong coupling to the adjacent antiferromagnetic layer. 

For studying the magnetodynamic properties conventional vector network analyzer (VNA)-FMR was used. The sample was placed face-down on a commercially available coplanar waveguide (CPW) and the scattering (S-)parameters measured by the VNA (Rohde\&Schwarz ZNB20). The dc magnetic field, provided by an electromagnet (GMW 3470), was probed by a Hall sensor. % (LakeShore FP-2X-250-TS15M-6)  

%\section{Exchange bias}

Modifying the thickness of the antiferromagnetic PtMn layer significantly influences the FMR spectra (see Fig.\,\ref{fig1}(b), straight lines). For PtMn thicknesses $t_\mathrm{PtMn} < 10$\,nm, only a single FMR peak is observed within the accessible magnetic field range. At $t_\mathrm{PtMn} = 10$\,nm, a second peak emerges, with the two peaks rapidly diverging as the PtMn thickness increases. This behavior suggests the presence of two coupled modes, likely originating from the interaction between the pinned and reference layers.
However, as the pinned layer becomes increasingly exchange-coupled to the antiferromagnet, this coupling could lead to a partial decoupling of the individual FMR modes. To investigate this possibility, we measured the FMR spectra of PtMn/CoFe bilayers across a similar PtMn thickness range (see Fig.\,\ref{fig1}(b), dashed lines). In this simpler system, only a single peak is observed, as only one ferromagnetic layer is present. Notably, this FMR peak exhibits a similar qualitative evolution as the dominant low-field peak in the pSAF, reinforcing the conclusion that the observed spectral behavior in both cases is governed by exchange-bias coupling to the antiferromagnet.

\begin{figure}
\includegraphics[width=10cm]{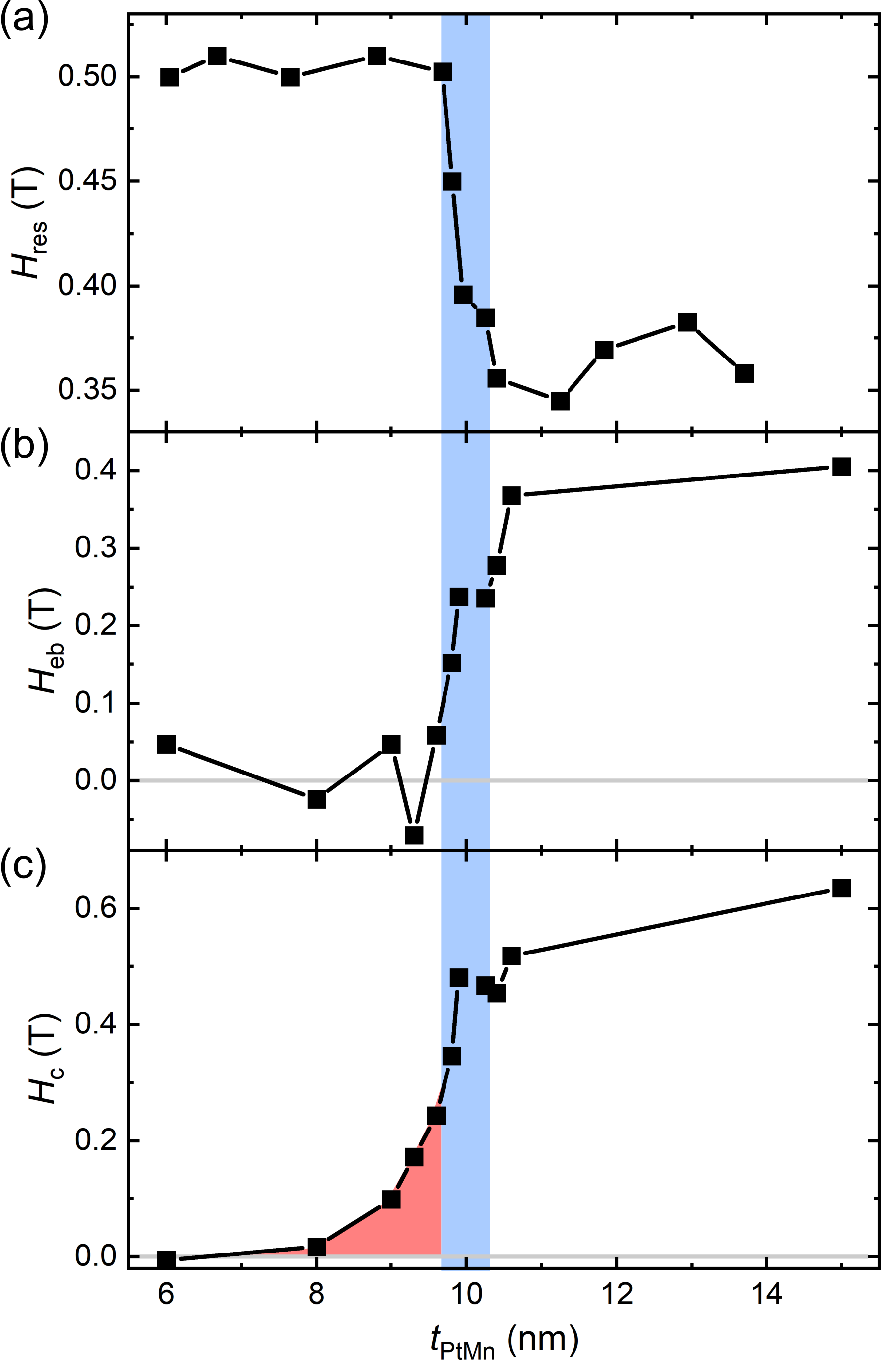}
\caption{\label{fig2} (a) Resonance field $H_\mathrm{res}$ as a function of the PtMn thickness for the low-field peak in (a). The increase of thickness leads to a abrupt decrease in $H_\mathrm{res}$ at arount 10\,nm. (b) $H_\mathrm{eb}$ and (c) $H_\mathrm{c}$ obtained from hysteresis measurements as function of PtMn thickness. }
\end{figure}

The magnetic resonance field $H_\mathrm{res}$, extracted from the FMR spectra, is plotted in Fig.\,\ref{fig2}(a) as a function of PtMn thickness $t_\mathrm{PtMn}$. As the thickness increases, $H_\mathrm{res}$ decreases sharply from approximately 500\,mT to 370\,mT. This transition occurs within a narrow thickness window of less than ±0.5\,nm, centered around 10\,nm. This abrupt drop in $H_\mathrm{res}$ indicates the onset of exchange bias.

The hysteresis loops of pSAFs with varying PtMn thicknesses are shown in Fig.\,\ref{fig1}(a). At a PtMn thickness of 6\,nm, the loops exhibit the characteristic symmetric scissor-like shape typical of RKKY-coupled ferromagnets. As the PtMn thickness increases, this symmetry is progressively lost due to the onset of exchange bias. The emergence of exchange bias is marked by the appearance of an additional hysteresis loop in the negative field branch, which becomes wider and shifts its center position with increasing thickness -- indicating changes in both the coercive field ($H_\mathrm{c}$) and the exchange bias field ($H_\mathrm{eb}$). 

The thickness dependence of $H_\mathrm{eb}$ and $H_\mathrm{c}$ is shown in Fig.\,\ref{fig2}(b) and (c), respectively. The exchange bias field $H_\mathrm{eb}$ -- defined as the center of the shifted hysteresis loop -- exhibits an abrupt onset near $t_\mathrm{PtMn} = 10$\,nm, closely mirroring the behavior of $H_\mathrm{res}$ (compare the shaded blue regions in Fig.\,\ref{fig2}(a) and (b)). In contrast, $H_\mathrm{c}$ begins to increase already below this critical thickness, as seen in the red-shaded region of Fig.\,\ref{fig2}(c). 

The same measurements have been repeated for the PtMn/CoFeB-bilayer (see Figures\,S1-S3 in the Supplemental Materials\,\cite{SM}). Figure\,S3(c) shows even more clearly the enhanced coercivity over the whole range of $t_\mathrm{PtMn}$, in particular before the onset of exchange bias. 

A likely origin of the early increase in coercivity is rotational anisotropy\cite{McMichael1998}. In this scenario, small regions of the antiferromagnet are coupled to the ferromagnet and partially dragged during magnetization reversal, but their volume is insufficient to fully pin the ferromagnet. This interaction contributes to enhanced coercivity even before the establishment of a stable exchange bias field.

\begin{figure}
\includegraphics[width=10cm]{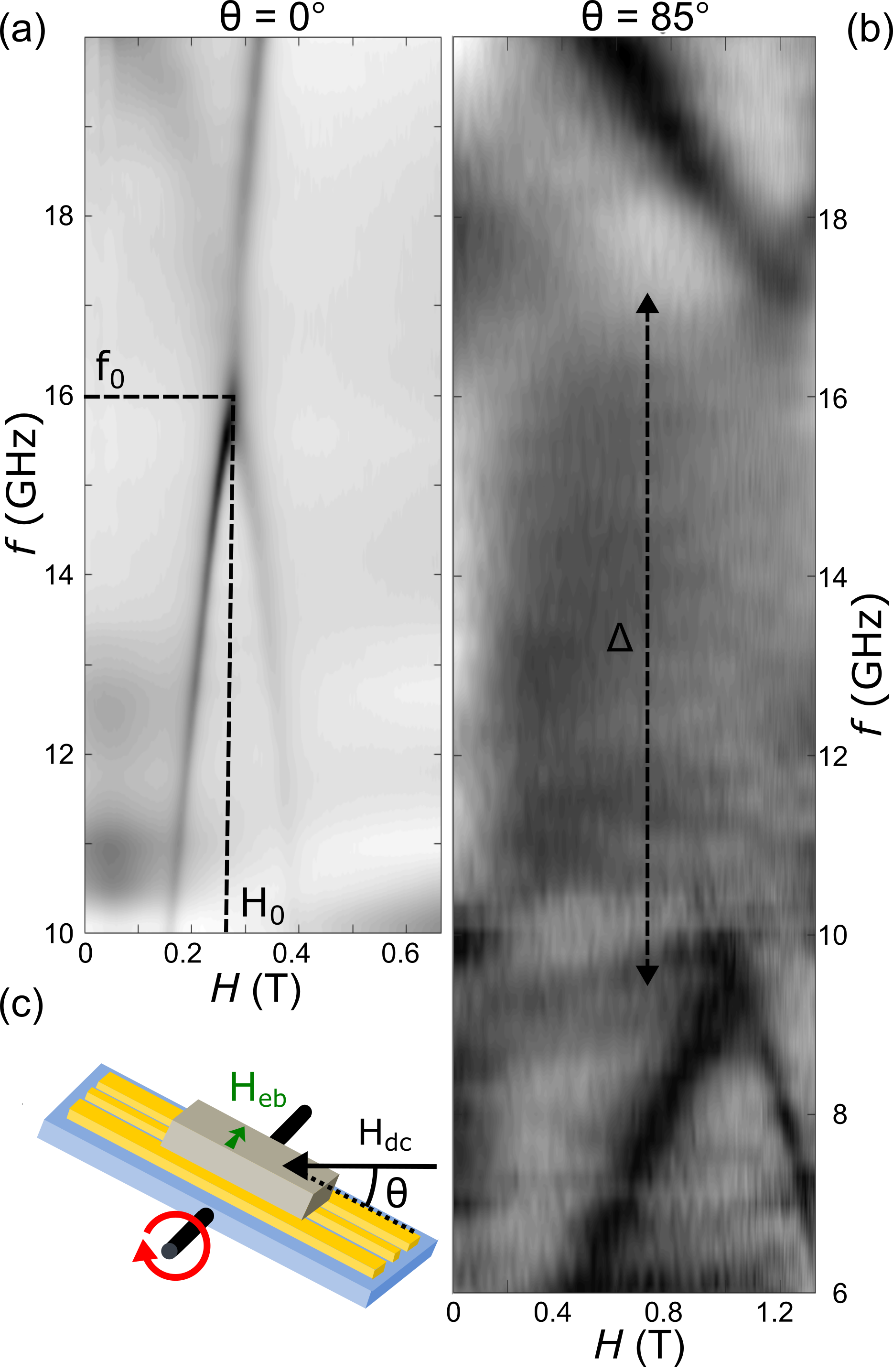}
\caption{\label{fig3} Spectral maps of a pSAF with $t_\mathrm{PtMn}=5.96$\,nm. The magnetic field $H_\mathrm{dc}$ was oriented at (a) $\theta$=0$^\circ$ and (b) $\theta$=85$^\circ$ with respect to the sample surface. (c) Schematic of the FMR waveguide and sample when tilted by $\theta$ with respect to $H_\mathrm{dc}$.} 
\end{figure}

%\section{Optical and acoustic FMR modes}

The precession of the ferromagnetic layers in the pSAF occurs in phase within the acoustic mode. This synchronized precession leads to a small net magnetic moment in the dynamical component, which interacts with the ac magnetic field. This interaction explains why the acoustic mode can be excited with relative ease.  

The optical mode lacks a net precessional magnetic moment because the magnetizations of the two ferromagnetic layers are 180° out of phase and cancel each other out. This cancellation occurs only in a perfectly symmetric system where both layers are magnetically equivalent. Introducing a slight asymmetry can enable excitation of the optical mode. In practice, this can be achieved by adjusting the thickness of one of the ferromagnetic layers\cite{Sud2023}, or by using different materials\cite{Ma2023}, although this breaks the magnetic compensation of the synthetic antiferromagnet (SAF).
Alternatively, the optical mode can be excited without sacrificing magnetic compensation by perturbing the precessional dynamics. For instance, rotating the waveguide azimuthally in the external magnetic field misaligns the ac and dc fields from their typical perpendicular configuration, thus facilitating excitation of the optical mode\cite{Sud2020,Shiota2020}.
In this work, asymmetry is introduced by pinning one of the ferromagnetic layers to an adjacent antiferromagnet, which modifies the magnetic anisotropy without altering the magnetization. As a result, the system remains magnetically compensated, meaning there is no net magnetic moment. This is a key advantage for technical applications—such as MRAM devices—where stray magnetic fields can degrade performance.

Figure \ref{fig3}(a) shows the spectral map of a pSAF with a PtMn thickness of 5.96\,nm, i.e., before the onset of exchange bias. Two intersecting dispersion branches are observed: one increases with applied field, while the other decreases. The upward-dispersing branch is more intense and exhibits a narrower linewidth, consistent with the acoustic mode. In contrast, the downward-dispersing branch corresponds to the optical mode, which is less intense due to its more challenging excitation conditions. Its broader linewidth is attributed to enhanced damping from the spin-pumping effect. This damping is absent in the acoustic mode, where spin pumping is symmetry-forbidden.

%\section{Magnon bandgap}

Having established well-defined optical and acoustic modes in the pSAF -- without modifying the intrinsic magnetic properties of the ferromagnetic layers or compromising the magnetically compensated nature -- we now explore the potential of such systems for magnon-magnon coupling. A key signature of mode hybridization is the emergence of a bandgap in the magnonic spectrum, indicating mixing between the acoustic and optical modes.
This bandgap typically becomes visible when the magnetic field is tilted from the in-plane to the out-of-plane direction. Figure \ref{fig3}(b) shows the spectral map at a tilt angle of $\theta = 85^\circ$, i.e., 5° away from full out-of-plane orientation. A clear splitting into upper and lower magnon branches is observed, marking the presence of a bandgap. The size of this gap, $\Delta$, is indicated by an arrow in Fig.\,\ref{fig3}(b), and the frequency at the original crossing point of the acoustic and optical modes, $f_0$, is 16\,GHz (see Fig.\,\ref{fig3}(a)). The coupling strength is defined as the ratio $\eta = \Delta / f_0$, which reaches a maximum value of 0.5 at $\theta = 85^\circ$.
Previously, values of $\eta$ ranging from 0.07 to 0.38 have been reported for magnetically compensated SAFs\,\cite{Sud2020,Shiota2020}. The coupling strength achieved in this work is the highest reported to date for a SAF system and approaches the threshold of the ultra-strong coupling regime, commonly defined as $\eta \geq 1$. 

\begin{figure}
\includegraphics[width=10cm]{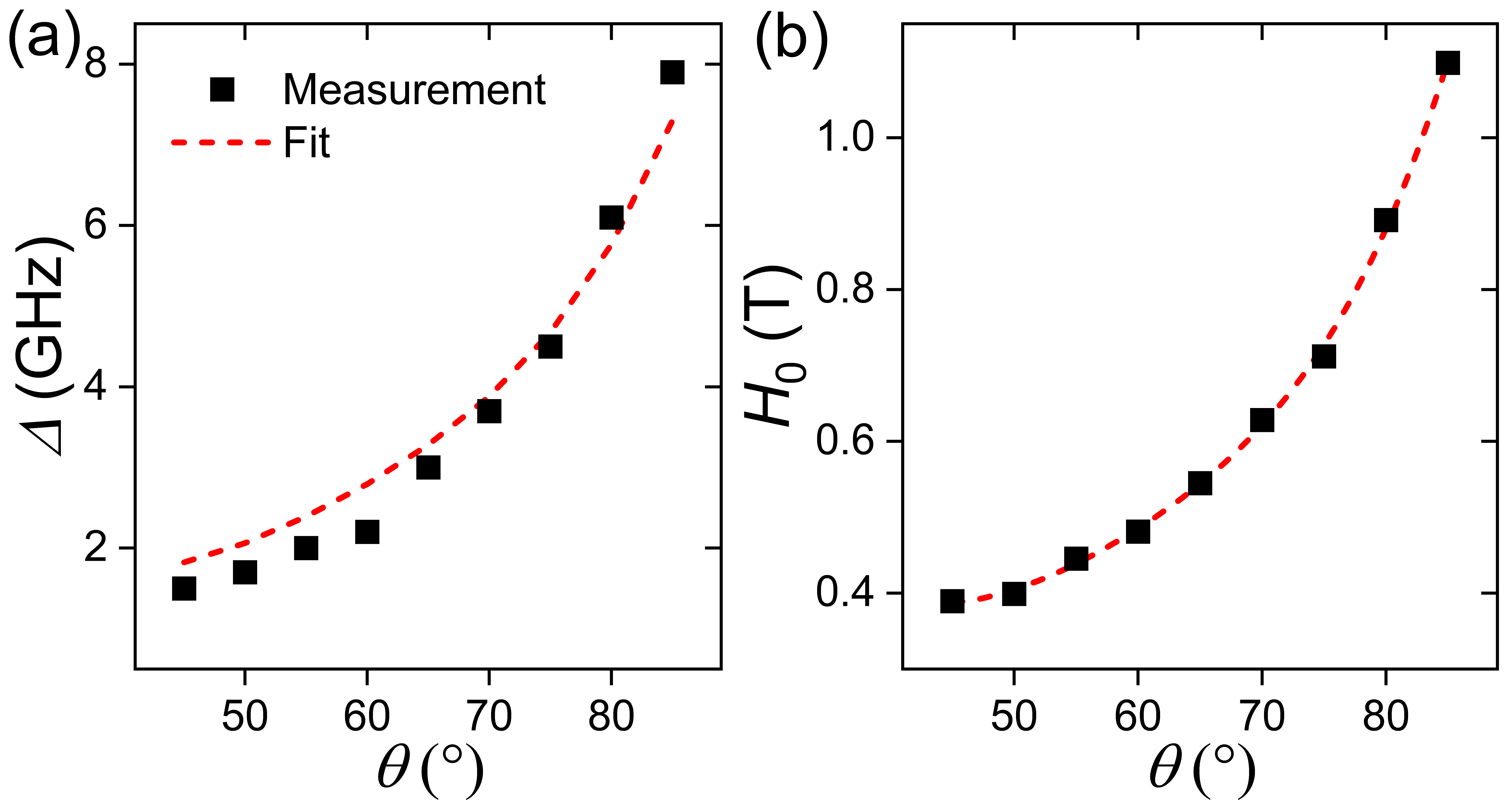}
\caption{\label{fig4} (a) Bandgap $\Delta$ and (b) resonance field $H_\mathrm{0}$ as a function of $\theta$. The red line in (a) is a fit of Eqn.\,\ref{gap} to the measurements.}
\end{figure}

Figure\,\ref{fig4}(a) and (b) show the bandgap\,$\Delta$ and the field position in which the intersection occurs, $H_0$. The corresponding spectral maps for each tilt angle $\theta$ can be found in the Supplemental Materials\,\cite{SM} (see Fig. S4). The bandgap $\Delta$ is given by

\begin{equation}
\Delta=\frac{1}{2\pi}\frac{\gamma H_\mathrm{ex}H_\mathrm{0}}{2H_\mathrm{s}+4H_\mathrm{ex}}  
\label{gap}
\end{equation}

with $H_\mathrm{0}$ the magnetic field position at the intersection, $H_\mathrm{s}=4\pi M_\mathrm{s}$ the saturation field and saturation magnetization, respectively, and $H_\mathrm{ex}$ the interlayer exchange coupling field\,\cite{Sud2020}. The angular dependence of the resonance field $H_\mathrm{0}$ is extracted from Fig.\,\ref{fig4}(b) using polynomial fitting. The saturation field $H_\mathrm{s} = 1.22$\,T is independently determined from magnetometry measurements. In the fitting procedure, the interlayer exchange field $H_\mathrm{ex}$ is treated as a free parameter and can be benchmarked against values obtained from other techniques to assess the validity of the model.
Figure\,\ref{fig4}(a) shows the resulting fit for $H_\mathrm{ex} = (0.091 \pm 0.012)$\,T. The model captures the overall shape of the angular dependence of the bandgap well but tends to overestimate its magnitude at angles $\leq 65^\circ$. Independent hysteresis measurements using vibrating sample magnetometry (VSM) yield a lower value of approximately 0.065\,T, while fitting the high-field FMR dispersion relation according to the method in Ref.\,\cite{Backes:2012} gives $H_\mathrm{ex} = (0.085 \pm 0.02)$\,T, which agrees well with the extracted value within the margin of error.

%\section{Conclusions}

In conclusion, we have demonstrated a method to simultaneously excite both optical and acoustic magnon modes in a synthetic antiferromagnet (SAF) without sacrificing its magnetically compensated nature -- an essential feature of SAFs and antiferromagnets more broadly. This is achieved by coupling one of the ferromagnetic layers to an antiferromagnet, forming a pinned SAF (pSAF). In this structure, we show that a magnonic band gap can be induced, depending on the strength of the coupling. Specifically, weak pinning -- before the onset of exchange bias -- allows efficient excitation of both modes and results in a large band gap. In contrast, strong pinning -- when the exchange bias is fully established -- suppresses the optical mode entirely, eliminating the gap (see Fig. S5 in the Supplemental Materials\,\cite{SM}).
This work is significant for several reasons. First, it enables strong magnon-magnon coupling using a standard VNA setup with conventional waveguides and without requiring cryogenic cooling. Second, the coupling strength reaches $\eta=0.5$, approaching the ultra-strong coupling regime and representing, to our knowledge, the highest value reported in thin-film SAFs to date. Third, the pSAF used here is integrated into a complete MRAM layer stack, fully compatible with standard silicon-based deposition and patterning techniques and easily scalable. Increasing the free layer thickness would re-enable established read-out and write mechanisms, such as those used in spin-transfer torque MRAM (STT-MRAM).
In summary, the pSAF platform constitutes a scalable, CMOS-compatible, tunable two-level system that operates under ambient conditions, offering strong potential for applications in quantum sensing and quantum information technologies.

\begin{acknowledgments}
We thank Diamond Light Source for access to the facilities of the Materials Characterisation Laboratory. We thank Dr. Jürgen Langer from Singulus Technologies for his assistance with sample preparation and Prof. Andy Kent from New York University for granting access to his laboratory for initial sample characterisation.

The data that support the findings of this study are available from the corresponding author upon reasonable request.
\end{acknowledgments}

%\appendix

%\nocite{*}
%

\clearpage

%\ifarXiv
%    \foreach \x in {1,...,\numbersupplementpages}
%    {
%        \includepdf[pages={\x}, fitpaper=true]{\supplementfilename}
%    }
%\fi

\ifarXiv
	

\section*{Supplementary material}

# Supplemental Material

# for

# Magnon-magnon coupling efficiency of $\eta$=0.5 in weakly pinned synthetic antiferromagnets

D. Backes[1]

[1]*Diamond Light Source, Harwell Science and Innovation Campus, Didcot OX11 0DE, United Kingdom*

(Dated: September 24, 2025)

## S1. MAGNETIC CHARACTERIZATION OF PTMN/COFE LAYERS

A control sample consisting of 3 Ta/40 Cu(N)/5 Ta/5-15 PtMn/2 CoFe/7 Ru on Si substrate with a 100nm SiO$_x$ layer on top was prepared to compare its magnetic properties with the pSAF sample. The absence of an exchange-coupled and free ferromagnetic layer facilitates the interpretation of the magnetometry data. Magnetic hysteresis measurements using a SQUID are shown in Fig. S1 for different PtMn thicknesses and the exchange bias field $H_{eb}$ and coercivity $H_c$ are shown in Fig. S3(b) and (c).

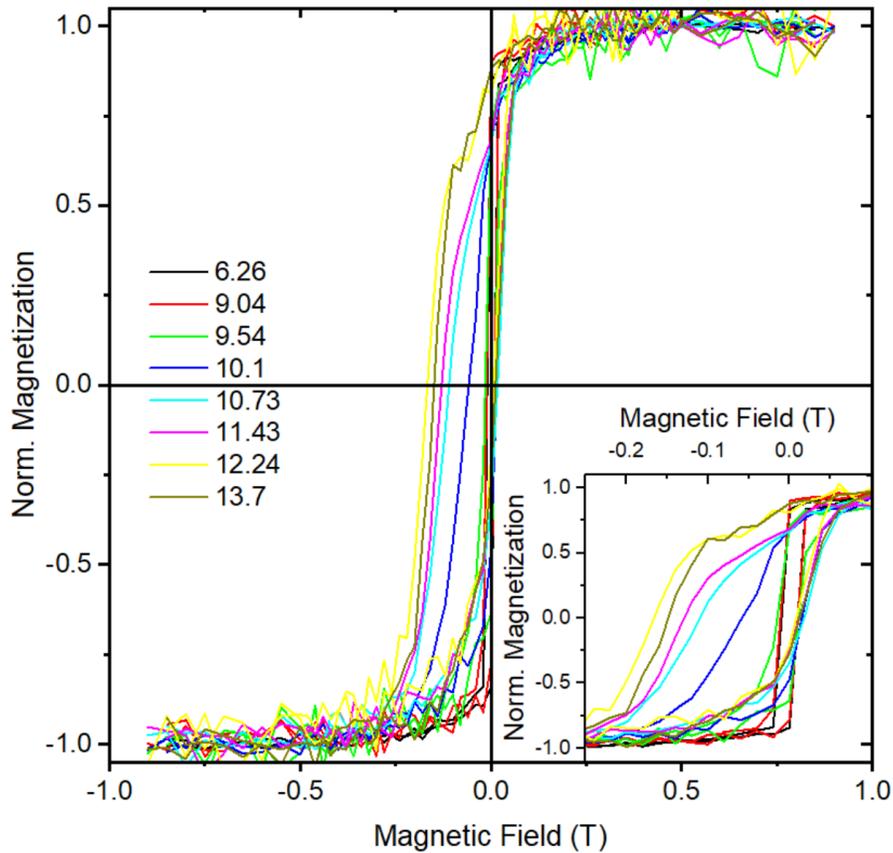

FIG. S1. Magnetic hysteresis measurements using a SQUID for various PtMn thicknesses. The inset shows a magnification of the exchange-biased loop.

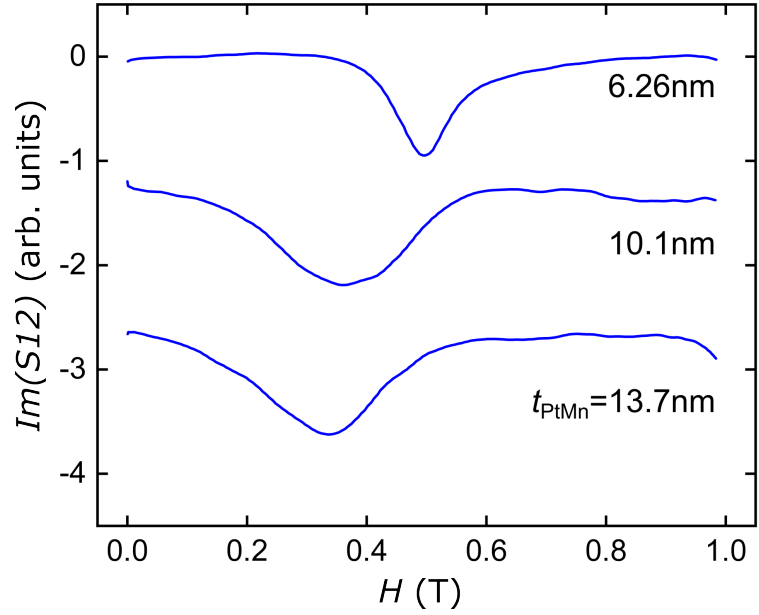

FIG. S2. FMR spectra for three different PtMn thicknesses

The FMR spectra for three different PtMn thicknesses are shown in Fig. S2. As the exchange-biased layer is the only ferromagnetic layer in the stack, the FMR peak can only originate from it. The spectra of the simple exchange bias sample are also shown in Fig. 1b in the main text. The resonance peak position $H_{res}$ as a function is shown in Fig. S3(a).

16

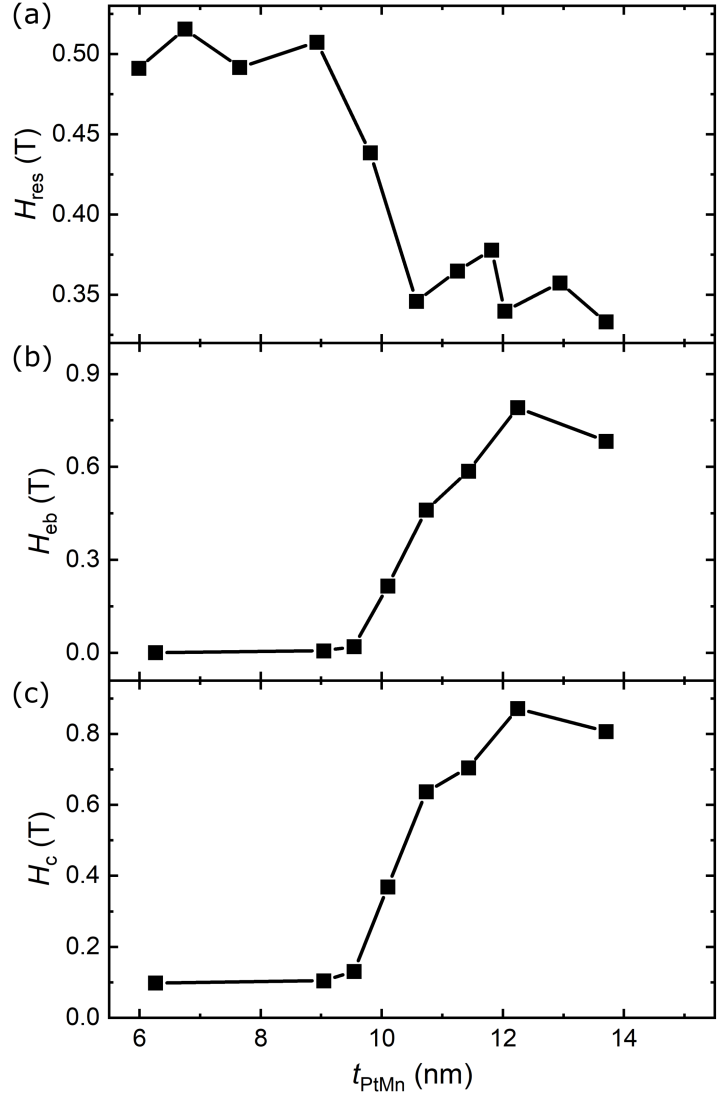

FIG. S3. FMR resonance peak position $H_{\text{res}}$, exchange bias field $H_{\text{eb}}$ and coercivity $H_{\text{c}}$ as a function of the PtMn layer thickness $t_{\text{PtMn}}$

The FMR resonance peak position $H_{\text{res}}$, exchange bias field $H_{\text{eb}}$ and coercivity $H_{\text{c}}$ are shown in Fig. S3. In all three cases a large change takes place at around $t_{\text{PtMn}} = 10$ nm, indicating that the peak shift originates from the onset of exchange bias around that thickness. A very similar effect was observed in the pSAF samples at the same thickness (see Figs. 1 and 2 in the main text). This is strong proof that the exchange bias causes the peak shift in the pSAF as well. Furthermore, it can be inferred that strong optical and acoustic modes can be observed in pSAF just before the full onset of exchange bias where the coercivity is slightly larger compared to an uncoupled CoFeB layer.

## S2. θ-DEPENDENT SPECTRAL MAPS

A series of FMR spectral maps taken at different orientations $\theta$ of the sample with respect to the magnetic field is shown in Fig. S4. The PtMn thickness was 5.96 nm. At $\theta = 0°$ the static magnetic field was aligned in the plane of the sample and coaxial with the waveguide. At $\theta = 90°$ the static magnetic field was aligned perpendicular to both sample and waveguide. Intermediate angles were positioned within the plane spanned by those two field orientations. At all orientations the static and ac magnetic field were perpendicular to each other. Both positive and negative fields are shown. The field was swept from negative to positive values and the sample rotated from $0°$ to $90°$. Changing the sweep or rotation direction had no influence on the result.

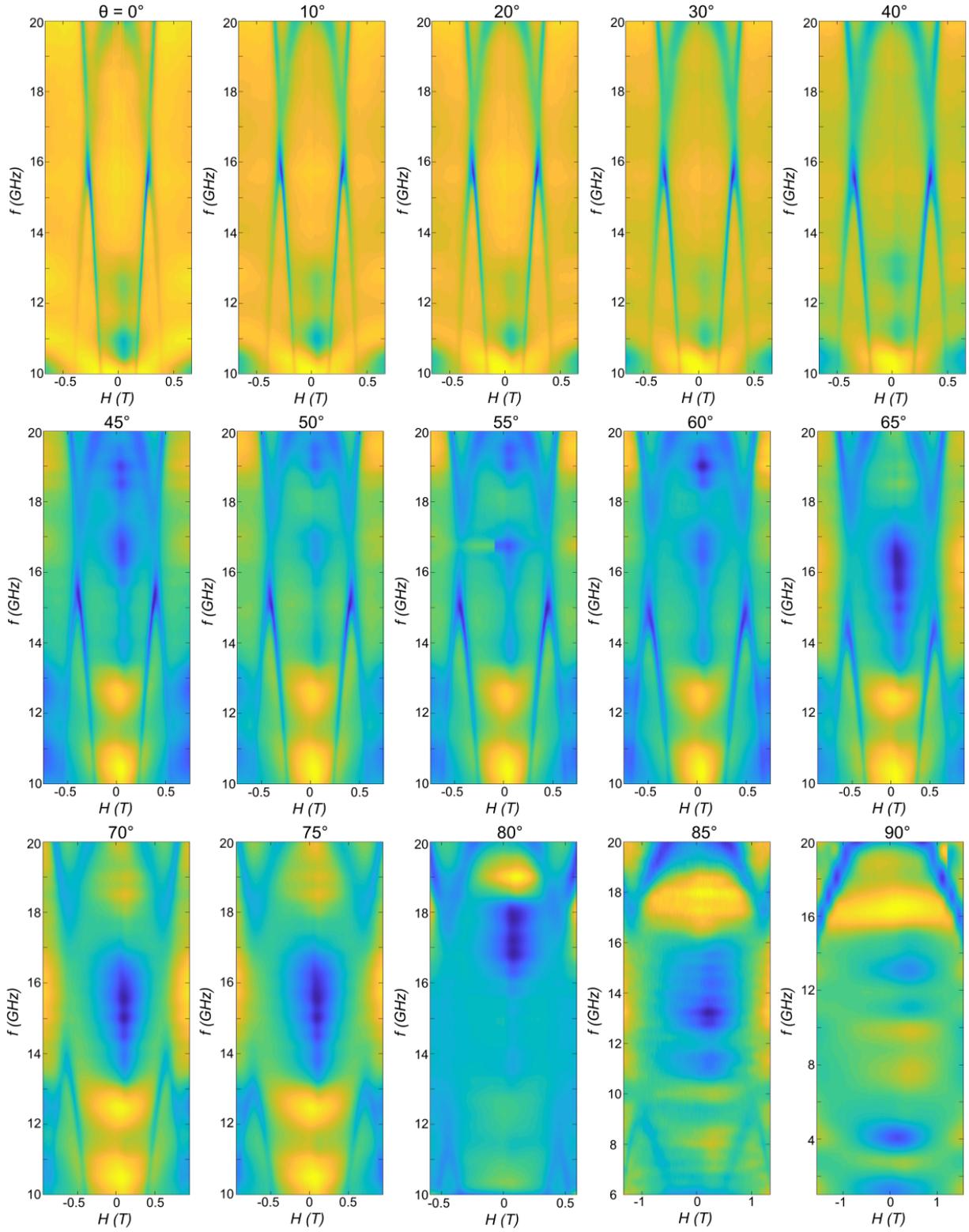

FIG. S4. FMR spectral maps taken at different tilt angles $\theta$ of the pSAF sample with respect to the magnetic field. The PtMn thickness was 5.96 nm.

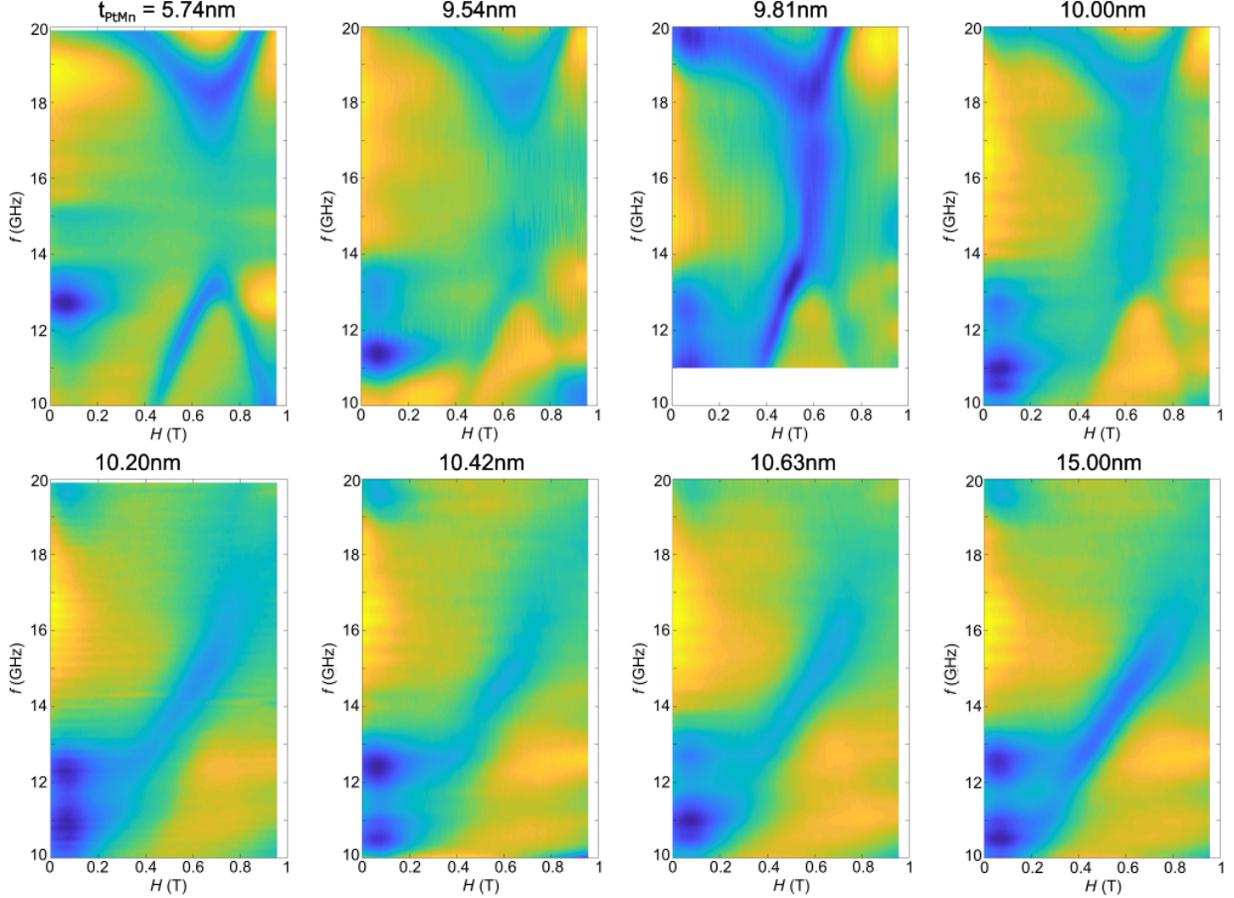

FIG. S5. FMR spectral maps taken at different thicknesses $t_{PtMn}$ of the antiferromagnetic PtMn of the pSAF sample. The samples was tilted by $\theta = 71°$.

## S3. $t_{PtMn}$-DEPENDENT SPECTRAL MAPS

The strength of pinning between the antiferromagnet and pinned layer was tuned by varying the thickness of the antiferromagnet, PtMn. The range below $t_{PtMn} = 10$ nm is characterized by weak pinning due to rotational anisotropy while above a strong unidirectional anisotropy caused by the exchange bias-effect dominates. Figure S5 shows the evolution of the FMR spectral maps with increasing $t_{PtMn}$. The tilt angle was kept at $\theta = 71°$. At $t_{PtMn} = 5.74$ nm a gap $\Delta$ of about 5 GHz is visible that remains intact up to $t_{PtMn} = 10$ nm. However, at $t_{PtMn} = 9.81$ nm spurious excitations are visible between the gap which like originate from inhomogeneity in the PtMn thickness. Above $t_{PtMn} = 10$ nm the optical mode disappears and the remaining mode is shifted towards higher frequencies.


\fi

\end{document}